\documentclass[8.5pt,twoside]{article}
\usepackage[super,sort&compress,comma]{natbib} 
\usepackage{times}
\usepackage{sectsty}
\usepackage{balance} 
\usepackage[text={11.3cm,20.4cm},centering]{geometry} 
\usepackage{graphicx} 
\usepackage{lastpage}
\usepackage[format=plain,justification=raggedright,singlelinecheck=false,font=small,labelfont=bf,labelsep=space]{caption} 
\usepackage{fancyhdr}
\usepackage	[center]	{subfigure}
\graphicspath{{./Graphs/}}

\begin{document}

\thispagestyle{plain}
\fancypagestyle{plain}{
\renewcommand{\headrulewidth}{1pt}}
\renewcommand{\thefootnote}{\fnsymbol{footnote}}
\renewcommand\footnoterule{\vspace*{1pt}%
\hrule width 11.3cm height 0.4pt \vspace*{5pt}} 
\setcounter{secnumdepth}{5}

\newcommand{\ang}{\mbox{\normalfont \mbox{\AA ngstr\" om}}}
\newcommand{\xray}{\mbox{X-ray}}
\newcommand{\mfp}{\mbox{MFPAD}}
\newcommand{\Fion}{F$^{+}$}
\newcommand{\Bion}{Br$^{+}$}

\makeatletter 
\renewcommand{\fnum@figure}{\textbf{Fig.~\thefigure~~}}
\def\subsubsection{\@startsection{subsubsection}{3}{10pt}{-1.25ex plus -1ex minus -.1ex}{0ex plus 0ex}{\normalsize\bf}} 
\def\paragraph{\@startsection{paragraph}{4}{10pt}{-1.25ex plus -1ex minus -.1ex}{0ex plus 0ex}{\normalsize\textit}} 
\renewcommand\@biblabel[1]{#1}            
\renewcommand\@makefntext[1]%
{\noindent\makebox[0pt][r]{\@thefnmark\,}#1}
\makeatother 
\sectionfont{\large}
\subsectionfont{\normalsize} 

\fancyfoot{}
\fancyfoot[RO]{\scriptsize{\sffamily{1--\pageref{LastPage} ~\textbar  \hspace{2pt}\thepage}}}
\fancyfoot[LE]{\scriptsize{\sffamily{\thepage~\textbar\hspace{3.3cm} 1--\pageref{LastPage}}}}
\fancyhead{}
\renewcommand{\headrulewidth}{1pt} 
\renewcommand{\footrulewidth}{1pt}
\setlength{\arrayrulewidth}{1pt}
\setlength{\columnsep}{6.5mm}
\setlength\bibsep{1pt}

\noindent\LARGE{\textbf{Imaging Molecular Structure through Femtosecond Photoelectron Diffraction on Aligned and Oriented Gas-Phase Molecules}}
\vspace{0.6cm}

\noindent\large{\textbf{Rebecca~Boll,$^{a,b,c}$ Arnaud~Rouz\'ee,$^{d,e}$ Marcus Adolph,$^{f}$ Denis~Anielski,$^{a,b,c}$ Andrew Aquila,$^{g,h}$ Sadia Bari,$^{h}$ C\'edric Bomme,$^{a}$ Christoph~Bostedt,$^{i}$ John~D.~Bozek,$^{i}$ Henry N. Chapman,$^{g,j,k}$  Lauge~Christensen,$^{l}$ Ryan~Coffee,$^{i}$ Niccola Coppola,$^{g,h}$ Sankar~De,$^{l,m}$ Piero~Decleva,$^{n}$ Sascha~W.~Epp,$^{a,b,c,o}$ Benjamin~Erk,$^{a,b,c}$ Frank Filsinger,$^{p}$ Lutz~Foucar,$^{b,q}$ Tais Gorkhover,$^{f}$ Lars Gumprecht,$^{g}$ Andr\'e H\"omke,$^{a,b}$ Lotte Holmegaard,$^{g,l}$, Per Johnsson,$^{r}$ Jens S. Kienitz,$^{g,k}$ Thomas Kierspel,$^{g,k}$  Faton~Krasniqi,$^{b,o,q}$ Kai-Uwe K\"uhnel,$^{c}$ Jochen Maurer,$^{l}$ Marc Messerschmidt,$^{i}$ Robert Moshammer,$^{b,c}$ Nele L. M. M\"uller,$^{g} $ Benedikt~Rudek,$^{b,c,s}$ Evgeny Savelyev,$^{a,v}$ Ilme Schlichting,$^{q}$ Carlo Schmidt,$^{b,c}$ Frank Scholz,$^{a}$  Sebastian~Schorb,$^{i}$ Joachim Schulz,$^{g,h}$ J\"orn Seltmann,$^{a}$ Mauro~Stener,$^{n}$ Stephan~Stern,$^{g,j}$ Simone~Techert,$^{a,b,u,v}$ Jan Th\o gersen,$^{l}$  Sebastian~Trippel,$^{g}$ Jens Viefhaus,$^{a}$ Marc~Vrakking,$^{d,e}$ Henrik~Stapelfeldt,$^{l}$ Jochen~K\"upper,$^{b,g,j,k}$ Joachim~Ullrich,$^{b,c,s}$ Artem~Rudenko,$^{b,c,t}$ and Daniel Rolles$^{\ast,a,b,q}$}}
\vspace{0.5cm}

\footnotetext{\textit{$^{\ast}$ Email: daniel.rolles@desy.de}}
\footnotetext{\textit{$^{a}$~Deutsches Elektronen-Synchrotron (DESY), 22607 Hamburg, Germany.}}
\footnotetext{\textit{$^{b}$~Max Planck Advanced Study Group at CFEL, 22607 Hamburg, Germany.}}
\footnotetext{\textit{$^{c}$~Max Planck Institute for Nuclear Physics, 69117 Heidelberg, Germany.}}
\footnotetext{\textit{$^{d}$~Max-Born-Institut, 12489 Berlin, Germany.}}
\footnotetext{\textit{$^{e}$~FOM-Institute AMOLF, 1098 XG Amsterdam, The Netherlands.}}
\footnotetext{\textit{$^{f}$~Technische Universit\"at Berlin, 10643 Berlin, Germany.}}
\footnotetext{\textit{$^{g}$~Center for Free-Electron Laser Science, DESY, 22607 Hamburg, Germany.}}
\footnotetext{\textit{$^{h}$~European XFEL GmbH, 22761 Hamburg, Germany.}} 
\footnotetext{\textit{$^{i}$~SLAC National Accelerator Laboratory, Menlo Park, California 94025, USA.}}
\footnotetext{\textit{$^{j}$~Department of Physics, University of Hamburg, 22761 Hamburg, Germany.}}
\footnotetext{\textit{$^{k}$~Center for Ultrafast Imaging, University of Hamburg, 22761 Hamburg, Germany.}}
\footnotetext{\textit{$^{l}$~Aarhus University, 8000 Aarhus C, Denmark.}}
\footnotetext{\textit{$^{m}$~Saha Institute of Nuclear Physics, 700064 Kolkata, India.}}
\footnotetext{\textit{$^{n}$~Dipartimento di Scienze Chimiche e Farmaceutiche, Universit\`a di Trieste, 34127 Trieste, Italy.}}
\footnotetext{\textit{$^{o}$~Max Planck Institute for Structural Dynamics, 22607 Hamburg, Germany.}}
\footnotetext{\textit{$^{p}$~Fritz-Haber-Institut der Max-Planck-Gesellschaft, 14195 Berlin, Germany.}}
\footnotetext{\textit{$^{q}$~Max Planck Institute for Medical Research, 69120 Heidelberg, Germany.}}
\footnotetext{\textit{$^{r}$~Department of Physics, Lund University, 22100 Lund, Sweden.}}
\footnotetext{\textit{$^{s}$~Physikalisch-Technische Bundesanstalt (PTB), 38116 Braunschweig, Germany.}}
\footnotetext{\textit{$^{t}$~J.R. MacDonald Laboratory, Kansas State University, Manhattan, Kansas 66506, USA.}}
\footnotetext{\textit{$^{u}$~Max Planck Institute for Biophysical Chemistry, 37077 G\"ottingen, Germany.}}
\footnotetext{\textit{$^{v}$~Institute of X-ray Physics, 37077 G\"ottingen University, Germany.}}

\noindent\textit{\small{\textbf{Received Xth, Accepted Xth \newline
First published on the web Xth}}}

\noindent \textbf{\small{DOI: }}
\vspace{0.6cm}

\section*{Abstract}

\noindent \normalsize{This paper gives an account of our progress towards performing femtosecond time-resolved photoelectron diffraction on gas-phase molecules in a pump-probe setup combining optical lasers and an X-ray Free-Electron Laser. We present results of two experiments aimed at measuring photoelectron angular distributions of laser-aligned 1-ethynyl-4-fluorobenzene (C$_8$H$_5$F) and dissociating, laser-aligned 1,4-dibromobenzene (C$_6$H$_4$Br$_2$) molecules and discuss them in the larger context of photoelectron diffraction on gas-phase molecules. We also show how the strong nanosecond laser pulse used for adiabatically laser-aligning the molecules influences the measured electron and ion spectra and angular distributions, and discuss how this may affect the outcome of future time-resolved photoelectron diffraction experiments.}
\vspace{0.5cm}

\section{Introduction}

\subsection{Motivation}

The prospect of studying chemical reactions with femtosecond resolution has been an inspiration for many experimental and theoretical investigations ever since the possibility of producing femtosecond light or electron pulses was first discussed.\cite{zewail_femtochemistry:_2000, chergui_electron_2009} 
Methods such as time-dependent mass spectrometry\cite{} and absorption spectroscopy \cite{chergui_picosecond_2010, pollard_analysis_1992} can provide information on the changes of the molecular structure that occur during chemical reactions by comparing the observed time-dependent signatures to theoretical predictions. More recently, methods aiming at imaging the structural changes more directly, for example by ultrafast \xray\cite{}  or electron diffraction\cite{chergui_electron_2009, sciaini_femtosecond_2011} were developed. In most cases however, their interpretation still heavily relies on comparison to theoretical models, and their time-resolution, in particular for the case of electron diffraction, has, to date, barely broken the one-picosecond mark.\cite{sciaini_femtosecond_2011, ihee_direct_2001, hensley_imaging_2012}

Free-Electron Lasers~(FELs) that produce intense, few-femtosecond light pulses in the vacuum ultraviolet~(VUV) and \xray~regime,\cite{ackermann_operation_2007, emma_first_2010, ishikawa_compact_2012,allaria_two-stage_2013} along with advances in the generation of (sub-)femtosecond pulses with laser-based high-harmonic generation~(HHG) sources\cite{popmintchev_attosecond_2010, corkum_attosecond_2007, agostini_physics_2004} and with relativistic electron guns,\cite{delsim-hashemi_charge_2013} have added new fuel to the long-standing vision of recording \emph{molecular movies} with \ang~spatial and femtosecond temporal resolution. Ideally, these movies would contain real-space images of the changing molecular structure that can be obtained without the necessity of comparison to theoretical modelling.\cite{krasniqi_imaging_2010}

In this article, we discuss how time-resolved photoelectron diffraction may be used to directly visualize ultrafast structural changes of gas-phase molecules, such as the formation of short-lived intermediate states during photodissociation or isomerization reactions. As an introduction, we discuss the relationship between molecular-frame photoelectron angular distributions and photoelectron diffraction in section 1.2. 
Section~2 briefly describes the experimental setup used to measure time-resolved photoelectron angular distributions of laser-aligned molecules at an FEL, and section~3 presents the results of these experiments. Here, we focus on data that has not been included in our previous publications\cite{boll_femtosecond_2013, rolles_femtosecond_2014} such as a comparison of ion time-of-flight spectra recorded at an FEL and at a synchrotron (section~3.1), effects of molecular orientation on photoelectron and fragment ion angular distributions (section~3.2), and the influence of both the alignment laser pulse and the femtosecond "pump" laser pulse on the photoelectrons and on the molecular photofragmentation process (sections~3.3 and~3.4, respectively). Our findings are summarized and conclusions for future time-resolved photoelectron diffraction experiments are drawn in section~4.

\subsection{Photoelectron diffraction and molecular-frame photoelectron angular distributions}

\begin{figure} [tb]
\centering
\includegraphics[width = 0.55\textwidth]{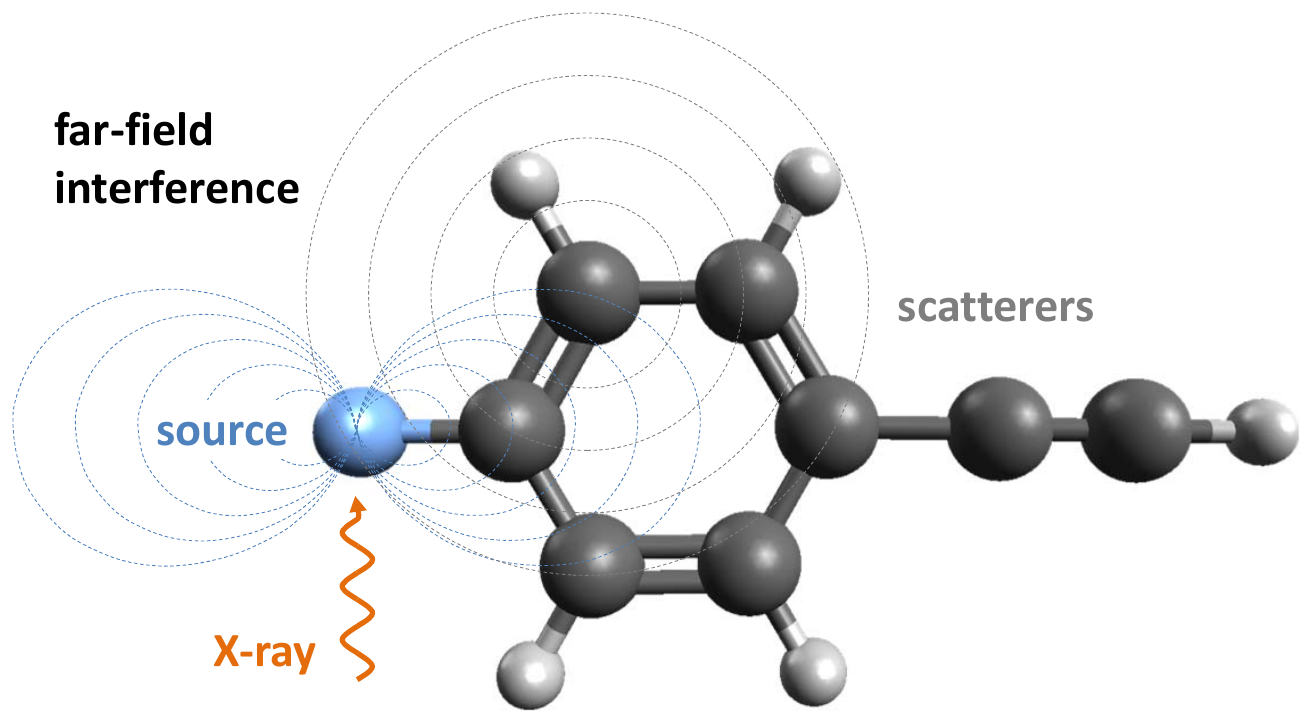}
\caption{\label{f_photoelectron}Schematic illustration of the photoelectron diffraction concept for a gas-phase C$_8$H$_5$F molecule: The emitted inner-shell photoelectron wave (blue), here created from the F($1s$) level by linearly polarized \xray s, scatters on neighboring atoms inside the molecule. The superposition of direct and scattered photoelectron waves, drawn here only for one of the neighboring carbon atoms, creates an interference pattern in the far field which contains information on the molecular structure.}
\end{figure}

The possibility to measure molecular-frame photoelectron angular distributions (\mfp s) of gas-phase molecules with electron-ion coincidence techniques developed in the 1990s\cite{golovin_photoionization_1992, shigemasa_angular_1995, heiser_demonstration_1997,downie_molecule-frame_1999, lafosse_vector_2000, dorner_cold_2000} led to a break-through in the study of molecular photoionization. Measurements of MFPADs allow, for example, the determination of photoionization matrix elements and phases \cite{motoki_complete_2002,gessner_4-1_2002} as well as investigations of core hole localization \cite{rolles_scattering_2005, schoffler_ultrafast_2008} and of the role of coherence and double-slit interferences in molecular photoemission.\cite{rolles_scattering_2005, rolles_isotope-induced_2005, akoury_simplest_2007} 
Extending the concept of photoelectron diffraction, which is a well-established method in solid state and surface physics,\cite{woodruff_adsorbate_1994, fadley_atomic-level_2008} to gas-phase molecules, it was realized early on that \mfp s of inner-shell electrons could also be interpreted in terms of diffraction.\cite{becker_photoelectron_2000, landers_photoelectron_2001, rolles_scattering_2005, zimmermann_localization_2008}
This opens up the possibility to obtain direct information on the geometric structure of the molecule from the photoelectron angular distribution, as illustrated schematically in Fig.~\ref{f_photoelectron} for the case of  F($1s$) inner-shell photoionization of a C$_8$H$_5$F molecule. Within the photoelectron diffraction model, the fluorine atom is considered as the source of photoelectrons that may scatter on the neighboring atoms in the molecule. The MFPAD is interpreted as the superposition of direct and scattered waves, creating an interference pattern on a detector in the far field, which contains structural information. This information is usually lost in gas-phase experiments on randomly oriented molecules because the diffraction pattern averages out when integrated over all molecular orientations. It can only be observed when the orientation of the molecule in the laboratory frame at the time of the electron emission is known.

In the surface physics community, scattering and diffraction of inner-shell photoelectrons is used, for example, to determine the geometry of molecules adsorbed on surfaces,\cite{fadley_study_1992, woodruff_adsorbate_1994, fadley_atomic-level_2008} thus providing insights into processes like catalytic reactions.\cite{} In contrast, the concept of photoelectron diffraction did not gain much interest in the gas-phase community, probably because far more precise methods, such as microwave spectroscopy, exist to determine the equilibrium structure of gas-phase molecules. Moreover, angle-resolved photoelectron-photoion coincidence measurements that have, so far, been used to determine the molecular orientation of gas-phase molecules are challenging and often time-consuming. 

With the availability of femtosecond VUV and \xray~sources that allow pump-probe studies involving inner-shell ionization, this situation is now changing. Time-resolved measurements of \mfp s and photoelectron diffraction of gas-phase molecules may offer information on ultrafast changes of molecular structure during chemical reactions which is difficult to obtain by other techniques.\cite{krasniqi_imaging_2010, berrah_non-linear_2010, ullrich_free-electron_2012, boll_femtosecond_2013,kazama_photoelectron_2013}

In this paper, we give an account of our experimental progress towards performing such femtosecond time-resolved experiments by combining optical lasers with VUV and soft \xray~FELs. The underlying idea is to first initiate a photochemical reaction with a "pump" laser pulse, and then to create an inner-shell photoelectron with an FEL pulse in order to image the molecules \textit{from within}. As a first step, we focus on measuring delay-dependent changes in the photoelectron angular distributions and on linking them to changes in the molecular geometry via comparison to density functional theory calculations.  The long-term goal is to employ the photoelectron diffraction concept in order to directly image molecular structure, for example by holographic reconstruction.\cite{krasniqi_imaging_2010}

\section{Experimental setup}

The experiments were performed at the Atomic, Molecular, and Optical Physics (AMO) beamline\cite{bostedt_ultra-fast_2013} of the Linac Coherent Light Source (LCLS)\cite{emma_first_2010} at SLAC National Accelerator Laboratory and at the Variable Polarization\footnote{At the time of the experiment, only circular polarization was available.} XUV Beamline P04\cite{viefhaus_variable_2013} of the synchrotron radiation source PETRA~III at DESY using the CFEL ASG Multi-Purpose (CAMP) endstation.\cite{struder_large-format_2010} The setup has been described in\cite{boll_femtosecond_2013, kupper_x-ray_2013, stern_stephans_2014} and, in detail, in \cite{rolles_femtosecond_2014}, and is only briefly summarized here. A beam of rotationally cold 1-ethynyl-4-fluorobenzene (C$_8$H$_5$F, pFAB) or 1,4-dibromobenzene (C$_6$H$_4$Br$_2$, DBB) molecules seeded in helium was created by supersonic expansion into vacuum and crossed with the \xray~beam inside a double-sided velocity map imaging (VMI) spectrometer. 

For the PETRA experiments, the molecular beam was operated continuously, and electrons and ions were detected using two microchannel plate (MCP) detectors equipped with Roentdek delay-line anodes, which record the time of flight and hit positions of multiple particles in coincidence. The amplified MCP and delay-line anode signals were processed by a hardware constant fraction discriminator and a multi-hit time-to-digital converter and then stored as a listmode event file. At the LCLS, a pulsed molecular beam was used, and electrons and ions were detected using MCP detectors with phosphor screens, that were read out for each FEL shot by 1-Megapixel CCD cameras. For time-of-flight measurements, the MCP signal traces were recorded for each FEL shot with an Acqiris DC282 digitizer. Processing of the single-shot CCD images, including a peak-finding algorithm, data sorting, and filtering on FEL machine parameters (photon energy and FEL pulse energy), was performed with the \mbox{CFEL-ASG} Software Suite (CASS).\cite{foucar_casscfel-asg_2012} The data shown here were taken during two LCLS experiments in 2010 (DBB) and in 2011 (pFAB) and during two PETRA experiments in 2013.

\subsection{Adiabatic laser alignment and orientation}

The determination of molecular orientation in an angle-resolved electron-ion coincidence experiment requires an ionization rate of less than one molecule per detection cycle in order to unambiguously correlate electrons and fragment ions. As the currently operating X-ray FELs have a maximum repetition rate of 120\,Hz, this technique yields very low count rates in FEL applications. An alternative approach to fix the molecular frame with respect to the laboratory frame is to actively align the molecules in space by using strong laser pulses.\cite{friedrich_polarization_1995, stapelfeldt_colloquium:_2003, stapelfeldt_alignment_2003, seideman_nonadiabatic_2006} This allows probing a whole ensemble of molecules with each FEL pulse, \cite{johnsson_field-free_2009, cryan_Auger_2010, cryan_molecular_2012, boll_femtosecond_2013, rolles_femtosecond_2014} thus dramatically increasing the achievable count rate.

At the LCLS, one- or three-dimensional adiabatic alignment was achieved by intersecting the molecular beam with pulses from a 1064\,nm, seeded neodymium-doped yttrium aluminum garnet~(YAG) laser with a pulse duration of 10-12\,ns and a pulse energy of 200-500\,mJ. A drilled mirror was used to collinearly propagate the YAG laser beam with the FEL beam, and the timing was set such that the FEL pulse arrived at the maximum of the YAG laser pulse, which corresponds to the maximum of the molecular alignment. \cite{sakai_controlling_1999} When using a linearly polarized YAG pulse, the molecules align such that their most-polarizable axis lies parallel to the laser polarization direction, which is the Br-Br axis in DBB and the F-C axis in pFAB. When using an elliptically polarized laser pulse, the second-most polarizable axis can be fixed in space as well. \cite{larsen_three_2000, nevo_laser-induced_2009} For the molecules used here, the plane of the benzene ring, which freely rotates for the case of one-dimensional alignment, is then also spatially confined. 

Moreover, one- or three-dimensional orientation can be achieved for polar molecules when an additional static electric field is present that has a vector component parallel to the polarization direction of the alignment laser field. \cite{friedrich_polarization_1995, holmegaard_laser-induced_2009, nevo_laser-induced_2009} In the presented data, the extraction field of the VMI spectrometer was used to define the direction of the fluorine atom in pFAB with respect to the electron and ion detectors.

For the experiments discussed here, the YAG laser operated at a repetition rate of 30\,Hz, and the LCLS at 60\,Hz in 2010 and at 120\,Hz in 2011, respectively. This allowed recording data for aligned and randomly oriented molecules concurrently. In 2011, the molecular beam was operated at 60\,Hz, such that background from residual gas could also be recorded concurrently. As shown in the following, this facilitates background subtraction substantially since long-term drifts were equally contained in each data subset.

\subsection{Three-color pump-probe experiments}

In order to initiate a structral change in the molecules via molecular fragmentation by strong-field ionization, an 800\,nm (1.55\,eV) titanium-sapphire~(TiSa) laser synchronized with the FEL was used in the 2010 LCLS experiments to pump the molecules before probing them with the FEL pulse. The TiSa laser beam was co-propagating with the YAG laser beam and the FEL beam, and the relative delay between FEL and TiSa pulses was varied using a delay stage. However, in 2010, the arrival time jitter between the TiSa pulse and the FEL pulse could not yet be corrected by \xray~optical cross-correlation,\cite{bionta_spectral_2011, schorb_x-rayoptical_2012, harmand_achieving_2013} and the temporal resolution of the pump-probe experiment was thus limited to 200-300\,fs. While this, among other technical difficulties, prevented the observation of delay-dependent changes in the photoelectron angular distribution, the experiment still demonstrated the feasibility of three-color pump-probe studies at an FEL.\cite{rouzee_towards_2013, rolles_femtosecond_2014} A subsequent pump-probe experiment at the LCLS in 2012 showed that with cross-correlation, the achievable temporal resolution is, at present, limited by the pulse durations of the TiSa laser and the FEL.\cite{Erk_ch3i_2014}

\section{Results and discussion}

\subsection{Fragmentation of pFAB molecules after inner-shell photoionization}

\begin{figure} [tb]
\centering
\includegraphics[width = 1.0\textwidth]{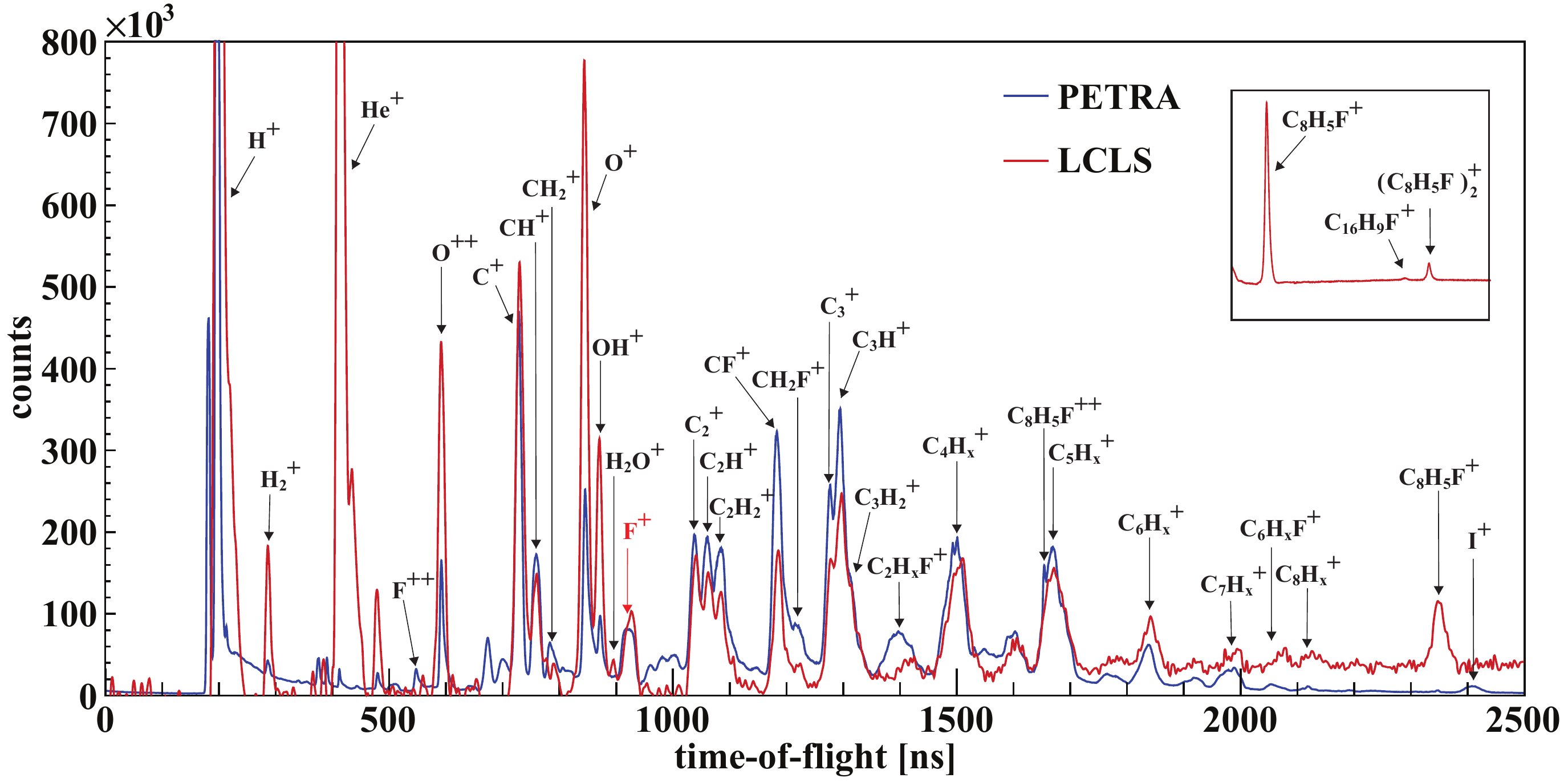}
\caption{\label{f_tof}Ion time-of-flight spectra of pFAB molecules obtained after photoionization with circularly polarized X-rays at a photon energy of 765\,eV  from the PETRA synchrotron radiation source (blue) compared to the spectrum obtained with linearly polarized \xray~pulses from the LCLS Free-Electron Laser at a photon energy of 742\,eV and 80\,fs pulse duration. The y-axis shows the total ions counts recorded in the PETRA spectrum, while the LCLS spectrum has been scaled and shifted such as to provide direct comparability with the synchrotron spectrum. The inset shows a zoom on the time-of-flight region with the parent ion and the dimer ion peak in the LCLS spectrum.} 
\end{figure}

In a polyatomic molecule, the core-hole created by inner-shell ionization typically decays within a few femtoseconds via single or multiple Auger decay. The resulting multiply charged molecular ion is usually not stable and subsequently dissociates into a variety of fragments. The charged fragments can be characterized by recording an ion time-of-flight (TOF) spectrum as shown in Fig.~\ref{f_tof} for the case of pFAB molecules ionized by X-rays from PETRA and the LCLS. The photon energies of 742 and 765\,eV lie approximately 50 and 73\,eV above the F($1s$) ionization threshold respectively (the F($1s$) binding energy in pFAB is assumed to be almost identical to the one in fluorobenzene, which is 692\,eV\cite{davis_x-ray_1972}). A large number of fragment ions from pFAB and residual gas can be identified. However, only a relatively small amount of \Fion~ions is produced despite the fact that, according to the photoabsorption cross-sections, every second \xray~photon that is absorbed ionizes the F($1s$) level.

Several ten eV above the F($1s$) ionization threshold, far beyond any potential shape resonances or other near-threshold phenomena, the fragmentation of pFAB can be considered to be rather insensitive to the exact photon energy. Therefore, the comparison of the ion TOF spectrum recorded using synchrotron radiation,
shown in blue in  Fig.~\ref{f_tof}, with the ion TOF spectrum obtained at the LCLS, shown in red, allows to identify the influence of possible multiphoton ionization that can occur due to the high intensity of the FEL pulse, as well as other influences stemming, for example, from the use of two different molecular beams in the PETRA and the LCLS experiments.\footnote{Note that the high number of ions detected per shot at the LCLS did not allow using a software constant fraction discriminator on the MCP trace to identify individual ion hits. Thus, the averaged MCP signal is shown which exhibits a slightly rising baseline towards higher times of flight.}

Overall, the two spectra are rather similar, showing that multiphoton processes are minor channels contributing to the overall fragmentation of the molecules.\footnote{The LCLS experiment was performed outside of the optimum focal position of the beamline, i.e.\,at an FEL spot size of approximately 30$\times$30\,$\mu$m$^2$, in order to reduce multi-photon ionization.} Besides a stronger contribution of water fragments in the LCLS spectrum, two main differences can be observed: A significantly larger He$^+$ peak in the spectrum recorded at the LCLS, and a relatively large amount of molecular parent ions in the LCLS spectrum, which are almost absent in the PETRA experiment. Whereas a continuous molecular beam with helium as a carrier gas at a relatively low backing pressure (few hundred millibars) was used at PETRA, the pulsed valve at the LCLS was operated with 50\,bar helium backing pressure resulting in a large number of helium atoms in the interaction zone. It also appears that for the expansion conditions in the LCLS experiment, a large amount of pFAB clusters was produced in the molecular beam, as indicated by the strong C$_8$H$_5$F$^+$ parent ion signal. This is further confirmed by the width of molecular parent ion peak, which indicates that the parent ions are produced with substantial kinetic energy, as well as by the singly charged pFAB dimer peak shown in the inset of Fig.~\ref{f_tof}. The presence of molecular clusters in the beam is particularly significant since these clusters are, most likely, not well aligned by the YAG pulse. Consequently, they produce a background of unaligned molecules in the ion and electron data recorded for aligned molecules at the LCLS. Unfortunately, the exact ratio of clusters to single molecules cannot be determined from the ion TOF spectra alone.

\begin{figure} [tb]
\centering
\subfigure{\includegraphics[width = 1.0\textwidth]{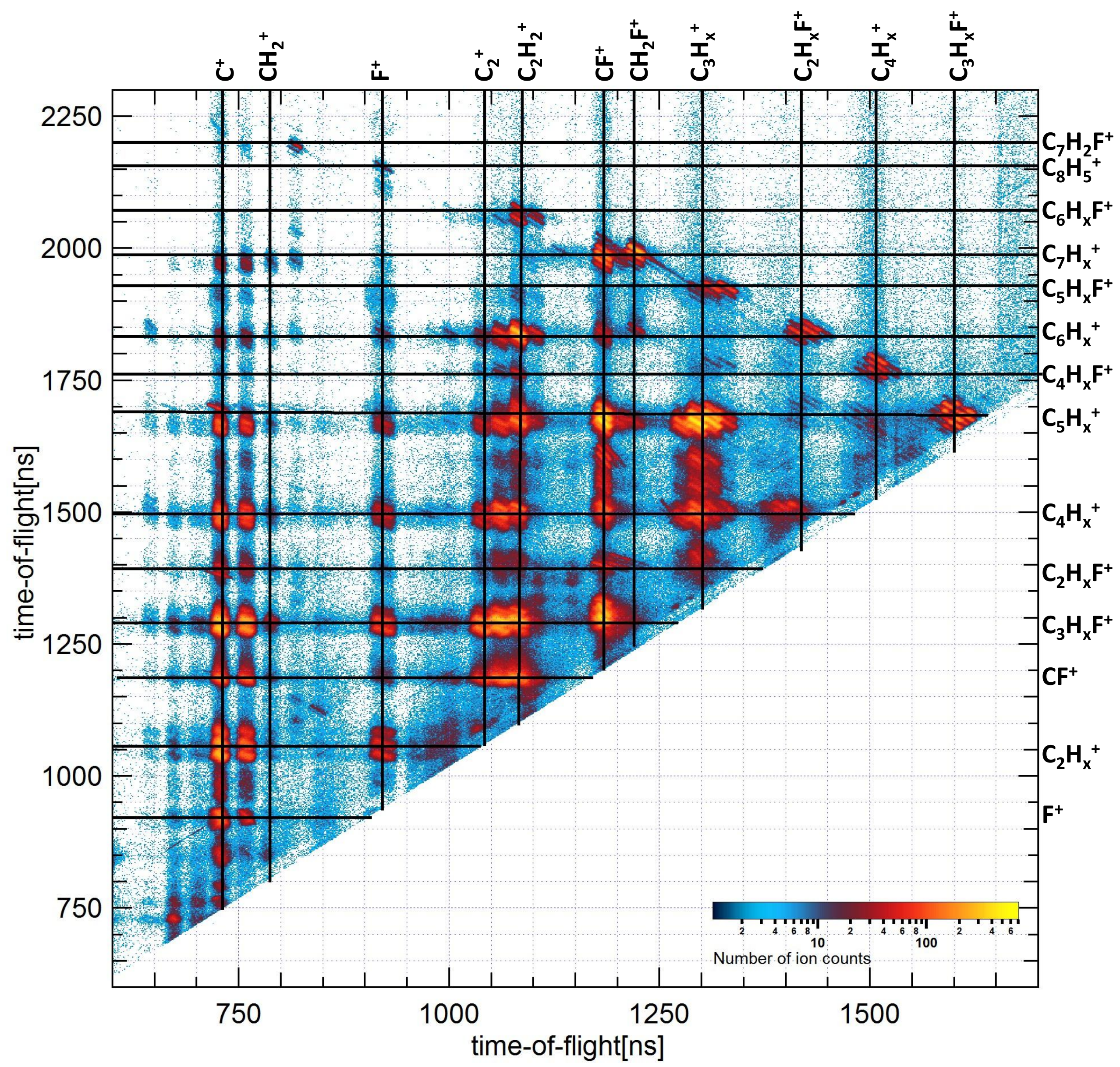}}
\caption{\label{PIPICO}Photoion-photoion coincidence (PIPICO) spectrum of pFAB molecules obtained after photoionization with circularly polarized X-rays from PETRA at a photon energy of 765\,eV.}
\end{figure}

Additional information on the fragmentation of pFAB molecules can be obtained when two or more charged fragments are recorded in coincidence, which can be represented in a photoion-photoion coincidence (PIPICO) map as shown in Fig.~\ref{PIPICO}. A large number of fragmentation channels can be identified, some of which correspond to the break-up of the pFAB molecules into two fragments, while at least a third fragment (either charged or neutral) must have been present in many of the break-up channels. Channels corresponding to the break-up into two charged fragments generally produce sharp diagonal lines in the PIPICO map as a result of momentum conservation. In contrast, when three or more charged fragments are created that each carry a significant amount of momentum, the corresponding line in the PIPICO map is more washed out. \cite{eland_dynamics_1987}  In  Fig.~\ref{PIPICO}, sharp PIPICO lines are observed for most of the break-up channels involving C$_5$H$_x$, C$_6$H$_x$, and C$_7$H$_x$ fragments, with the exception of C$_5$H$_x$-C$_2$H$_x$ ($x$ denotes varying numbers of H atoms). Most channels involving an \Fion~exhibit rather washed out lines, suggesting that these mostly stem from a break-up into at least three charged fragments, each carrying a significant amount of momentum. We note that this does not bode well for using \Fion~ions to determine the orientation of the F-C axis in an angle-resolved photoelectron-ion coincidence experiment. In the following, however, the emphasis shall not be put on further interpretations of the wealth of information that can be extracted from the momentum-resolved coincidence data but rather on the effects of the alignment laser on the electron and ion images and spectra recorded at the LCLS.

\subsection{Molecular alignment and orientation}

\begin{figure} [tb]
\centering
\subfigure[without YAG]{
\label{f_off_bgr_sub}	
\includegraphics[width = 0.22\textwidth]{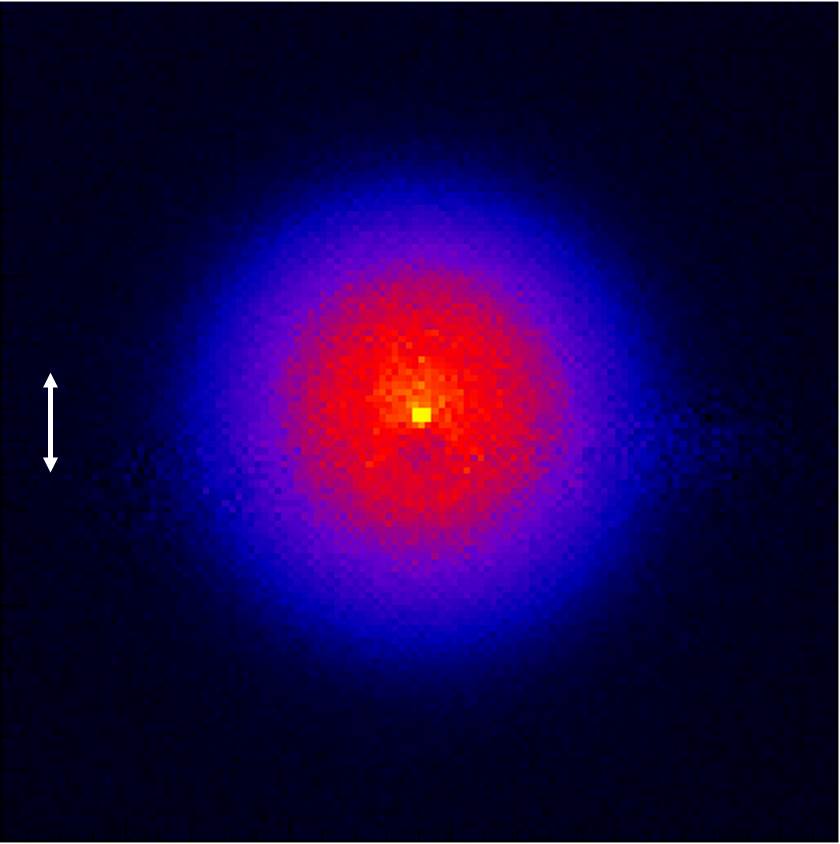}}
\subfigure[with YAG]{
\label{f_on_clust_sub}	
\includegraphics[width = 0.22\textwidth]{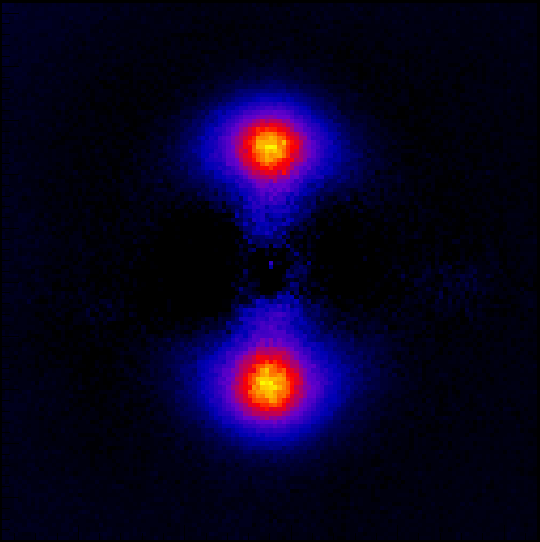}}
\subfigure[YAG at +45$^\circ$]{
\label{f_ori_up}		
\includegraphics[width = 0.22\textwidth]{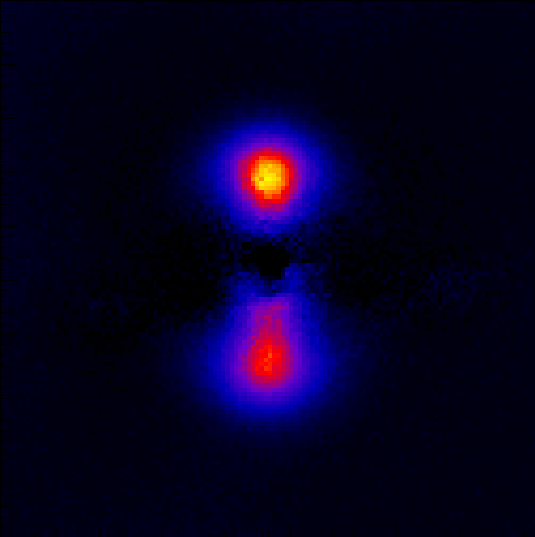}}
\subfigure[YAG at -45$^\circ$]{
\label{f_ori_down}	
\includegraphics[width = 0.22\textwidth]{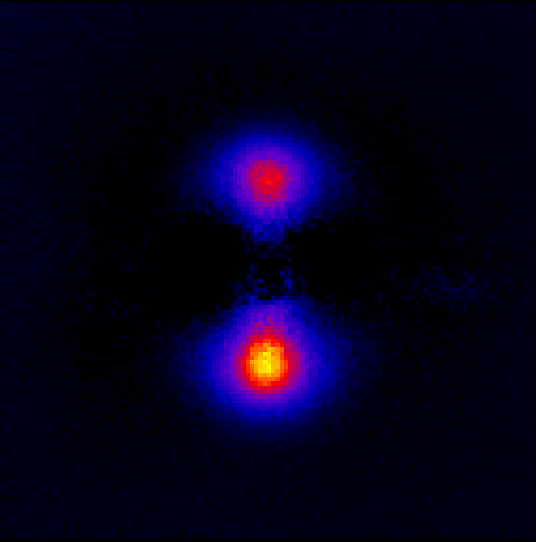}}
\subfigure{\includegraphics[width = 0.0435\textwidth]{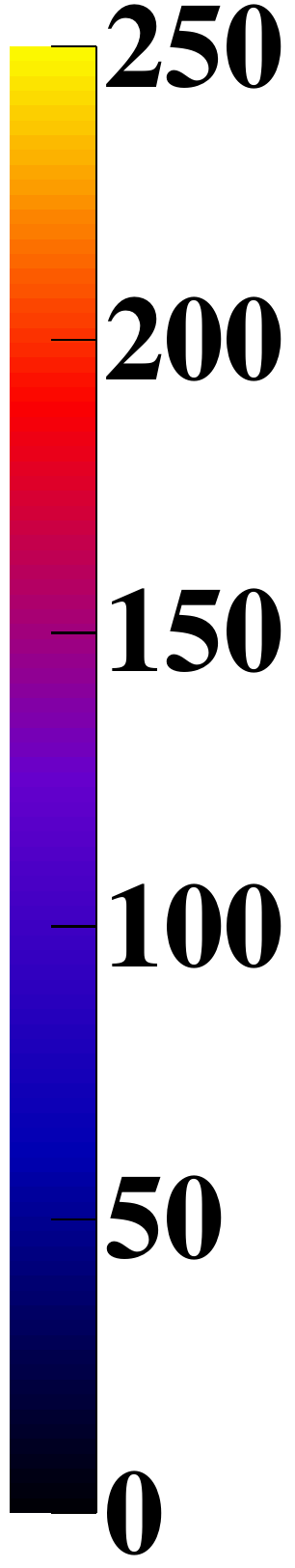}}
\subfigure[same as (a)]{
\label{3d_yag_off}			
\includegraphics[width = 0.468\textwidth]{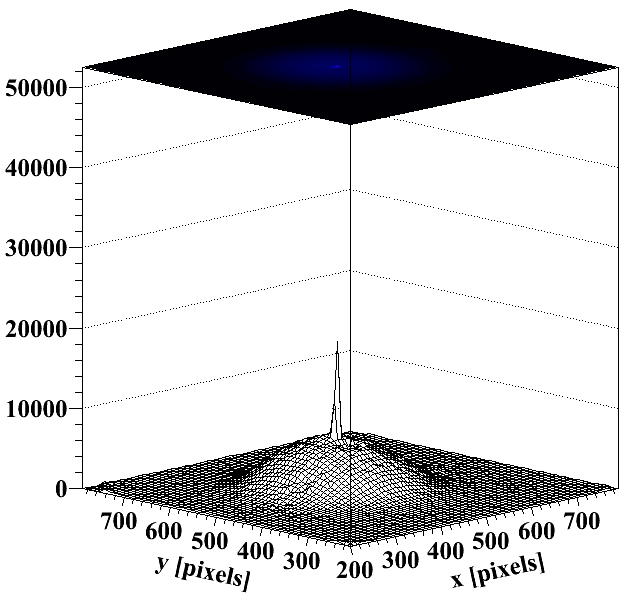}}
\subfigure[same as (b) but without subtraction]{
\label{3d_yag_on}			
\includegraphics[width = 0.468\textwidth]{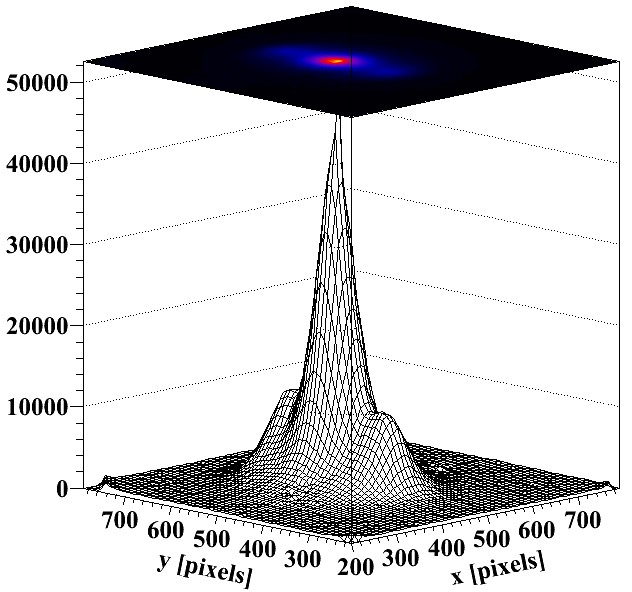}}

\caption{\label{f_detectors_on}\Fion~ion images recorded at the LCLS for ionization of pFAB molecules with linearly polarized X-rays at 723\,eV photon energy; (a) without the YAG laser pulse, (b) with the YAG laser pulse linearly polarized parallel to the FEL polarization, and (c) and (d) with the YAG polarization rotated out of the detector plane by +45$^\circ$ and -45$^\circ$, respectively. The polarization direction of the X-rays is indicated by the arrow in panel (a). The images were obtained by using a peak-finding algorithm on the single-shot CCD camera images. (f) and (g) show the same ion images as (a) and (b) but without subtraction of the low-energy \Fion~ions (see text).}
\end{figure}

In order to characterize the degree of alignment and orientation induced in the beam of pFAB molecules by the combination of the YAG laser pulse and the static electric field of the VMI spectrometer, the emission direction of the \Fion~ions can be used as a marker, assuming that they are emitted along the direction of the F-C axis. Fig.~\ref{f_detectors_on} shows the \Fion~ion images recorded at the LCLS for ionization of pFAB molecules with linearly polarized \xray s at a photon energy of 723\,eV, with and without the YAG laser pulses and for different directions of the YAG pulse polarization axis. The ion detector was gated by fast switching of the high voltage such that only hits in the time-of-flight interval corresponding to the arrival time of the \Fion~ions were detected. However, when operating the spectrometer in velocity map imaging mode, the signal of \Fion~ions (mass of 19\,amu) could not be fully separated from the signal of H$_{2}$O$^{+}$ ions (mass of 18\,amu), and a contribution from the ionization of residual water in the vacuum chamber was contained in the detector image.  Since this background was continuously recorded, it could be subtracted accurately, resulting in the images in Fig.~\ref{f_detectors_on}. The sharp dot in the center of Fig.~\ref{f_off_bgr_sub} corresponds to water that is present in the molecular beam, which was, hence, not removed by the background subtraction. 

Without the YAG pulse, the \Fion~hits are distributed isotropically, see Fig.~\ref{f_off_bgr_sub}, reflecting the random orientation of the F-C axis and the fact that the photoionization probability at this photon energy is almost independent of the molecular orientation with respect to the polarization direction of the \xray s, making the FEL an almost ideal probe for the molecular alignment. If linearly polarized YAG laser pulses are present, the \Fion~ions are emitted preferentially along the polarization of the YAG pulse, as seen in Fig.~\ref{f_on_clust_sub}, indicating a strong angular confinement of the F-C axis in the pFAB molecules at the time of the ionization by the FEL pulse.

Because of the above-mentioned pFAB clusters that were present in the molecular beam, the \Fion~images recorded with the YAG pulse present contain an additional contribution of isotropically distributed \Fion~ions with lower kinetic energies that stem from clusters, which are not aligned by the YAG pulses but which are fragmented if the YAG pulse is present, as discussed further in section 3.4. This contribution, which can clearly be seen in Fig.~\ref{3d_yag_on}, was fitted by a two-dimensional Lorentz distribution and subtracted from the ion images recorded with the YAG pulses in order to accurately determine the degree of molecular alignment. Only the resulting distribution of \Fion~ions from aligned pFAB molecules is shown in Figs.~\ref{f_on_clust_sub} to~\ref{f_ori_down}. The achieved degree of molecular alignment can be quantified by the ensemble-averaged expectation value of $\cos^2 \theta_{2D}$, where $\theta_{2D}$ is the angle between the projection of the \Fion~ion momentum vector on the detector plane and the polarization axis of the YAG laser pulse. It can be calculated from the integrated ion detector image as

\begin{equation}
\langle \cos^2 \theta_{2D}\rangle = \frac{ \sum_{i, j} I(R_{i}, \theta_{2D,j})\,\cos^2 \theta_{2D,j}}{\sum_{i, j} I(R_{i}, \theta_{2D,j})}
\label{eq_cos}
\end{equation}

where $I$ is the number of counts at a certain radius $R_{i}$, measured from the center of the distribution, and at a certain angle $\theta_{2D,j}$. For Fig.~\ref{f_on_clust_sub}, the resulting value is $\langle \cos^2 \theta_{2D}\rangle$\,=\,0.89. When integrating the two-dimensional distribution in Fig.~\ref{f_on_clust_sub} over R and fitting the resulting ion angular distribution with a Gaussian, this corresponds to a FWHM of 47$^\circ$.

When the polarization direction of the YAG pulses is rotated such that it does not lie perpendicular to the spectrometer axis, the extraction field of the VMI spectrometer is no longer perpendicular to the YAG polarization and thus induces orientation of the pFAB molecules. \cite{holmegaard_laser-induced_2009, nevo_laser-induced_2009}
The permanent dipole moment of the molecule is directed along the F-C axis from the F atom ("negative end") to the benzene ring ("positive end"). In our geometry, this means that the fluorine atom preferentially points away from the ion detector. Therefore, when the polarization direction of the YAG laser is turned by +45$^\circ$ or -45$^\circ$ with respect to the detector plane, the \Fion~ion images show an asymmetry, as can be seen in Fig.~\ref{f_ori_up} and~\ref{f_ori_down}. To the best of our knowledge, this is the first realization of mixed-field molecular orientation at an FEL. The degree of molecular orientation can be quantified by the ratio

\begin{equation}
\Delta N = \frac{N(F_{up}^+)}{N(F^+)} 
\label{eq_ori}
\end{equation}

where $N(F^+)$ is the integral of the complete detector image and $N(F_{up}^+)$ is the integral in the upper half of the detector. \cite{holmegaard_laser-induced_2009} This results in $\Delta N = 0.61$ and $\Delta N = 0.39$ for Figs.~\ref{f_ori_up} and~\ref{f_ori_down} respectively.

\subsection{Photoelectron angular distributions of aligned and oriented molecules}

\begin{figure} [tb]
\centering
\subfigure[FEL, pFAB]{
\label{f_el_off}						
\includegraphics[width = 0.24\textwidth]{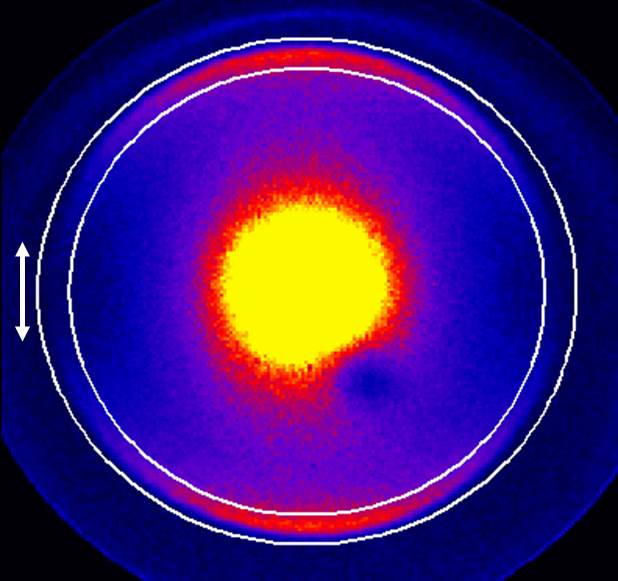}}
\subfigure[FEL, YAG, pFAB]{
\label{f_el_on}						
\includegraphics[width = 0.24\textwidth]{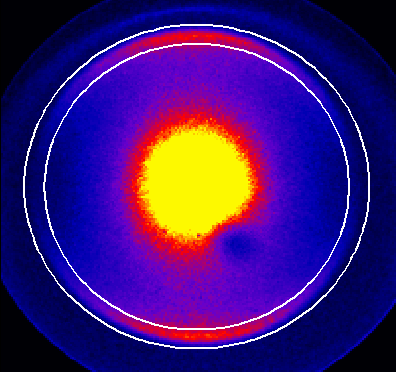}}
\subfigure{\includegraphics[width = 0.043\textwidth]{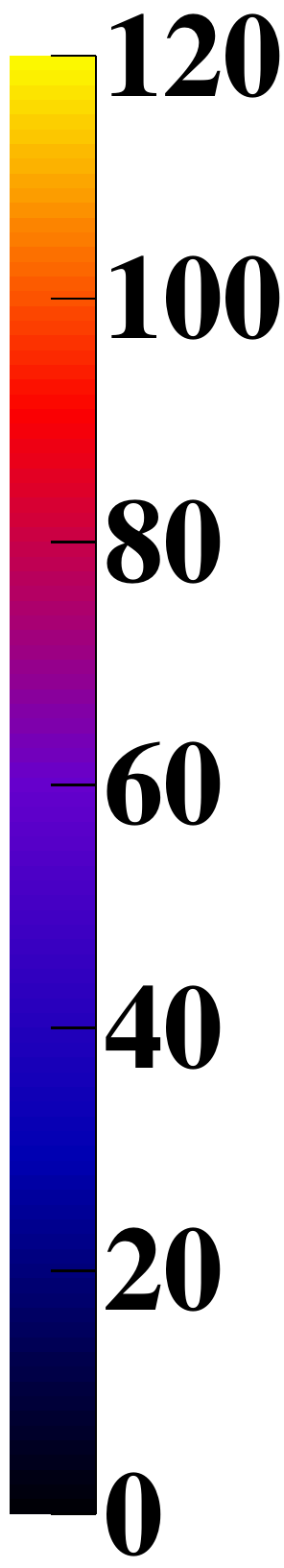}}
\subfigure[(b) minus (a)]{
\label{f_el_on_m_off}			
\includegraphics[width = 0.24\textwidth]{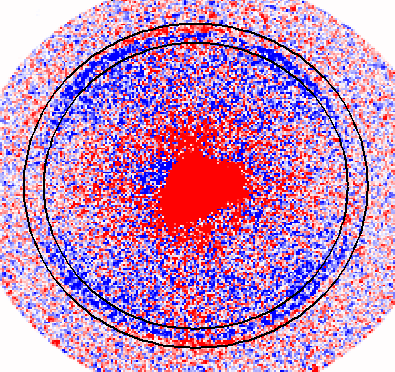}}

\subfigure[inversion of (a)]{
\label{f_el_off_inv}				
\includegraphics[width = 0.24\textwidth]{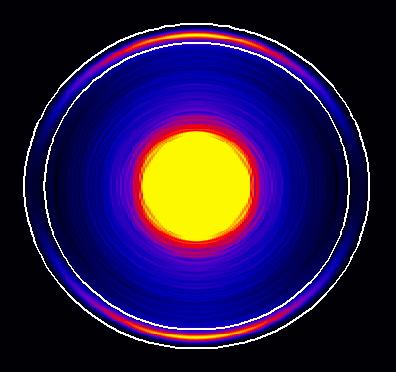}}
\subfigure[inversion of (b)]{
\label{f_el_on_inv}				
\includegraphics[width = 0.24\textwidth]{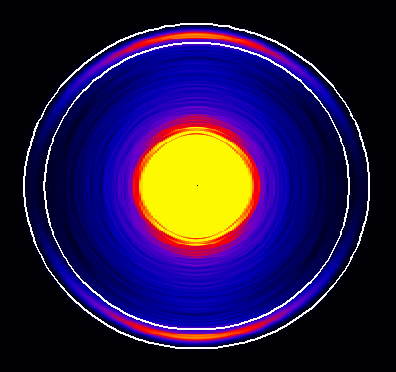}}
\subfigure[(f) minus (e)]{
\label{f_el_on_m_off_inv}	
\includegraphics[width = 0.24\textwidth]{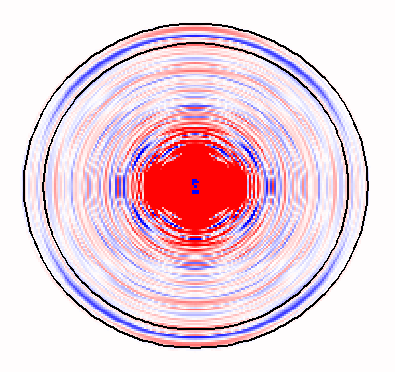}}
\caption{\label{f_el_det}Electron images from the ionization of randomly oriented and one-dimensionally aligned pFAB molecules by linearly polarized \xray s with 742\,eV photon energy. The polarization directions of the FEL and YAG pulses are parallel and indicated by the arrow. The top row shows the 2-D momentum images, the bottom row the inverted images obtained by applying the pBasex code.\cite{garcia_two-dimensional_2004} The top and bottom right panels show the difference between the images recorded with and without YAG pulses. In the difference plots, red corresponds to positive values, blue to negative values. }
\end{figure}

Simultaneously to the ion imaging, electrons are imaged on the other side of the velocity map imaging spectrometer, such that the photoelectron angular distributions can be determined. Figure~\ref{f_el_off} shows the integrated electron detector image obtained by using a peak-finding algorithm on the single-shot CCD camera images for randomly oriented pFAB molecules ionized by LCLS pulses at a photon energy of 742\,eV, resulting in F($1s$) photoelectrons of 51\,eV kinetic energy. The F($1s$) photoline is marked by the white circles. It shows the pronounced angular anisotropy expected for single-photon ionization of an $s$-orbital. In addition to the F($1s$) photoelectrons, a strong electron signal is observed in the center of the image, corresponding to electrons with lower kinetic energy. These electrons are most likely created by multi-electron processes such as Auger cascades, shake-up or shake-off, and inelastic scattering of photoelectrons or Auger electrons inside the molecule. High-energy electrons created from \mbox{C($1s$)} and valence ionization as well as fluorine and carbon KLL-Auger electrons have kinetic energies of \mbox{$>$240\,eV}, and are thus collected only in a small solid angle for the chosen spectrometer voltages and appear as a small, almost flat background.

The plots in the bottom row of Fig.~\ref{f_el_det} show the inverted electron images after applying the pBasex algorithm.\cite{garcia_two-dimensional_2004} The algorithm fits the electron angular distribution by an expansion in Legendre polynomials, which is a valid description of the angular distribution for the case of a cylindrically symmetric system such as one-dimensionally aligned molecules with the axis of alignment parallel to the detector plane. It is then possible to retrieve the full three-dimensional distribution from the experimentally recorded two-dimensional projections. The resulting images in the bottom row show a cut through the three-dimensional electron distribution in the detector plane. 

When comparing the electron images recorded with and without the YAG laser pulses in Figs.~\ref{f_el_off} and~\ref{f_el_on} or the inverted images in Figs.~\ref{f_el_off_inv} and~\ref{f_el_on_inv}, only small differences can be seen in the angular distribution of the F($1s$) photoelectrons. This can be explained by contributions from the unaligned molecular clusters to the electron signal, as well as by the averaging over different alignments of the molecular axis, which is confined to the YAG laser polarization axis only within a Gaussian of 47$^\circ$ FWHM. Despite the rather high degree of alignment of $\langle \cos^2 \theta_{2D}\rangle$\,=\,0.89, this averaging smears out possible interference structures, and the photoelectron angular distribution therefore looks very similar to the one for randomly oriented molecules. 

Plotting the difference between the images recorded with and without the alignment laser visually enhances the effect of the molecular alignment. An increase of the photoelectron intensity along the polarization direction of the YAG pulses and a decrease at 45$^\circ$ to it is clearly visible in Figs.~\ref{f_el_on_m_off} and \ref{f_el_on_m_off_inv}. This corresponds to a narrowing of the photoelectron angular distribution for aligned molecules as compared to randomly oriented molecules. We note that there is a radial dependence of this effect even \emph{within} the region of the F($1s$) photoline. We tentatively attribute this to the creation of sidebands of the main photoline due to "above-threshold" absorbtion of YAG photons by the photoelectrons \cite{meyer_two-colour_2010}, as described in more details in section~\ref{YAG}. Moreover, an increase of intensity in the center of the image is found when the YAG pulses are present, which we interpret as additional low-energy electrons created by the interaction of the YAG laser pulse with excited molecular fragments, as also explained in section~\ref{YAG}.

\begin{figure} [tb]
\centering
\subfigure[YAG at 0$^\circ$]{
\label{f_51}	
\includegraphics[width = 0.28\textwidth]{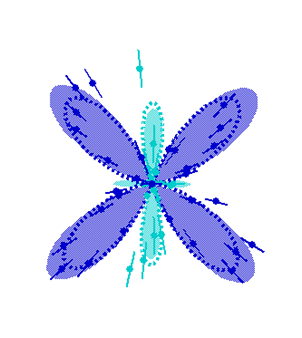}}
\subfigure[YAG at +45$^\circ$]{
\label{f_ori_up_el}		
\includegraphics[width = 0.28\textwidth]{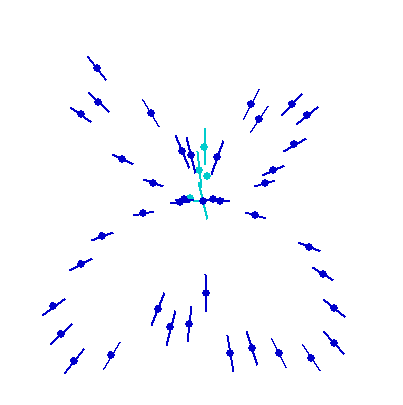}}
\subfigure[YAG at -45$^\circ$]{
\label{f_ori_down_el}	
\includegraphics[width = 0.28\textwidth]{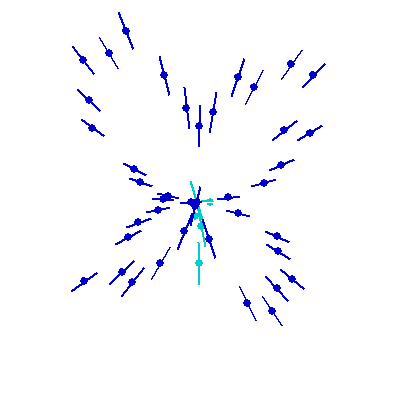}}
\caption{\label{f_el}Fluorine ($1s$) photoelectron angular distribution differences ($\Delta$PADs), shown as polar plots, for ionization of aligned (a) and oriented (b,c) pFAB molecules recorded for a photoelectron kinetic energy of 51\,eV. Positive differences are plotted in cyan, negative differences in blue. The data points are obtained by radial integration in the region of interest in Fig.~\ref{f_el_on_m_off}. The shaded areas in (a) are obtained from the inverted data in Fig.~\ref{f_el_on_m_off_inv}. Also shown in (a) as a dotted line is the calculated difference obtained from density functional theory.\cite{boll_femtosecond_2013}}
\end{figure}

A more quantitative analysis is possible when radially integrating the difference images over the region of interest containing the F($1s$) photoline, as defined by the circles in Figs.~\ref{f_el_on_m_off} and~\ref{f_el_on_m_off_inv}. The resulting photoelectron angular distribution differences ($\Delta$PADs)\cite{boll_femtosecond_2013} are shown in Fig.~\ref{f_51} as polar plots. The $\Delta$PADs obtained from both, the raw projection and the inverted image, agree well within the statistical uncertainties. The experimental data also agree very well with the results of DFT calculations. Further details on the DFT calculations and additional data for other photon energies are presented in \cite{boll_femtosecond_2013}.

Establishing the connection between the shape of the $\Delta$PADs and the molecular structure without comparison to theory is not straightforward for electrons with kinetic energies of only a few tens of~eV, since a direct reconstruction of the molecular geometry in a holographic sense\cite{krasniqi_imaging_2010} is not possible. However, the link of the $\Delta$PAD to the molecular geometry becomes clearer when the molecules are oriented in space instead of only being aligned. The resulting $\Delta$PADs for opposite molecular orientations are shown in Figs.~\ref{f_ori_up_el} and~\ref{f_ori_down_el}. The distribution is clearly mirrored when the fluorine atom points in opposite directions, as seen in the corresponding ion images in Figs.~\ref{f_ori_up} and~\ref{f_ori_down}. This clearly demonstrates the sensitivity of the photoelectron angular distribution to the molecular frame. An inversion of the VMI image for the case of oriented molecules can not be performed as the cylindrical symmetry is broken when the molecular axis is no longer parallel to the detector surface, thus only raw data are shown in this case.

\subsection{Effects of the alignment laser}
\label{YAG}

For the above discussion of the photoelectron angular distributions, it has been implicitly assumed that the alignment laser has no other effect besides fixing the molecular axes in space. Although it has been verified experimentally that the YAG pulse alone does not ionize the molecules, one has to keep in mind that in adiabatic alignment, the laser pulse is present \emph{during} and \emph{after} the \xray~pulse, which means that the ionization as well as all secondary processes happen in the presence of a strong laser field with a field strength on the order of 10$^{11}$\,W/cm$^{2}$. In this section, we will discuss some experimental evidences for resulting two-color effects.

\begin{figure} [tb]
\centering
\subfigure{\includegraphics[width = 1.0\textwidth]{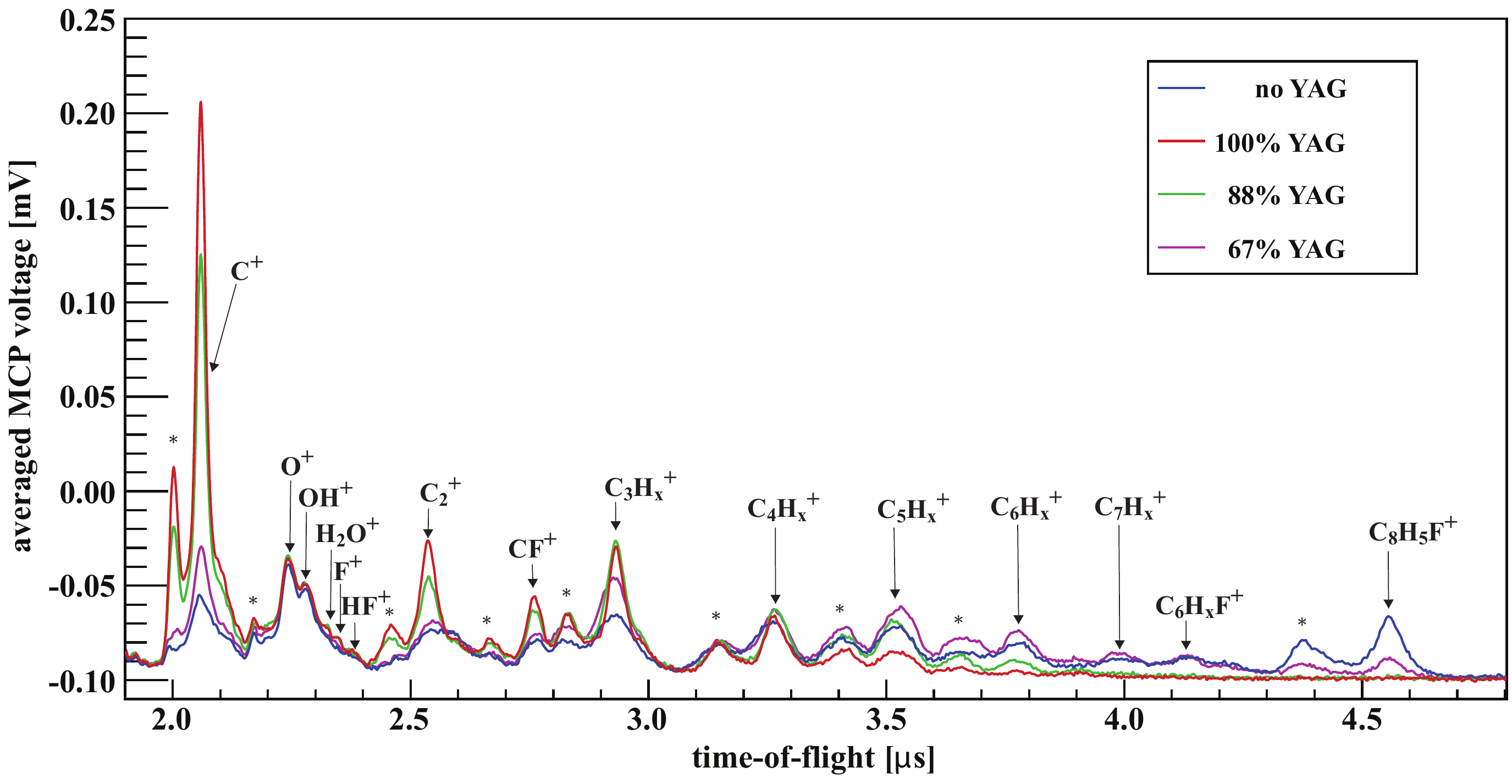}}
\caption{\label{f_yag_scan}Ion time-of-flight spectra of pFAB after ionization by 727\,eV \xray s from the LCLS in the presence of YAG laser pulses with different intensities. The polarization of the YAG pulses was parallel to the X-ray polarization direction and the full YAG intensity was about 5\,$\times$\,10$^{11}$\,W/cm$^{2}$. Contrary to the ion TOF spectra shown in Fig.~\ref{f_tof}, these spectra were measured in VMI focussing conditions, which results in a decreased time-of-flight resolution. Moreover, for the extraction voltages chosen here, secondary electrons created on the mesh that terminated the ion drift region resulted in additional peaks in the spectrum which are marked by asterisks.}
\end{figure}

The influence of the alignment laser on the fragmentation of pFAB molecules after inner-shell ionization was investigated by recording ion time-of-flight spectra at a photon energy of 727\,eV for different YAG pulse intensities. When comparing these spectra shown in Fig.~\ref{f_yag_scan}, it is obvious that the YAG pulses indeed influence the molecular fragmentation. Most notably, the largest ionic fragments, including the broad parent ion peak, are strongly suppressed or disappear completely when the YAG pulses are present, while the yield of smaller fragments increases. 

A possible explanation for this observation could be that the heavy fragments are produced in excited electronic states. Such excited fragments may occur due to shake-up processes during the photoionization or as intermediates during the following Auger decay, as suggested previously when interpreting HHG-pump infrared-probe experiments on small molecules \cite{sandhu_observing_2008, zhou_probing_2012} and FEL-pump optical-probe experiments on xenon atoms~\cite{krikunova_time-resolved_2009}. Either the photon energy or the intensity of the YAG pulse may thus be sufficient to dissociate or ionize these excited states with a single or a few photons, thereby producing smaller fragments. This is supported by the fact that some smaller fragments, namely C$_3$H$^+$, CF$^+$, C$_2^+$, and especially C$^+$, increase in yield when the YAG pulse is present. We note that the ions with the largest masses, most notably the C$_8$H$_5$F$^+$ parent ion peak, are produced mainly by \xray~ionization of pFAB clusters, and we cannot conclude from the present data if the post-dissociation or post-ionization by the YAG pulses affects these cluster fragments more strongly than the fragments stemming from individual molecules. 

\begin{figure} [tb]
\centering
\subfigure[]{\label{f_yag_el}\includegraphics[width = 0.27\textwidth]{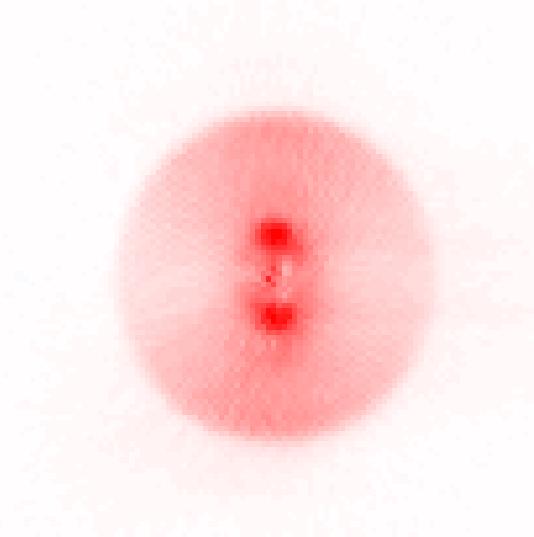}}
\subfigure[]{\label{f_spectrum}\includegraphics[width = 0.67\textwidth]{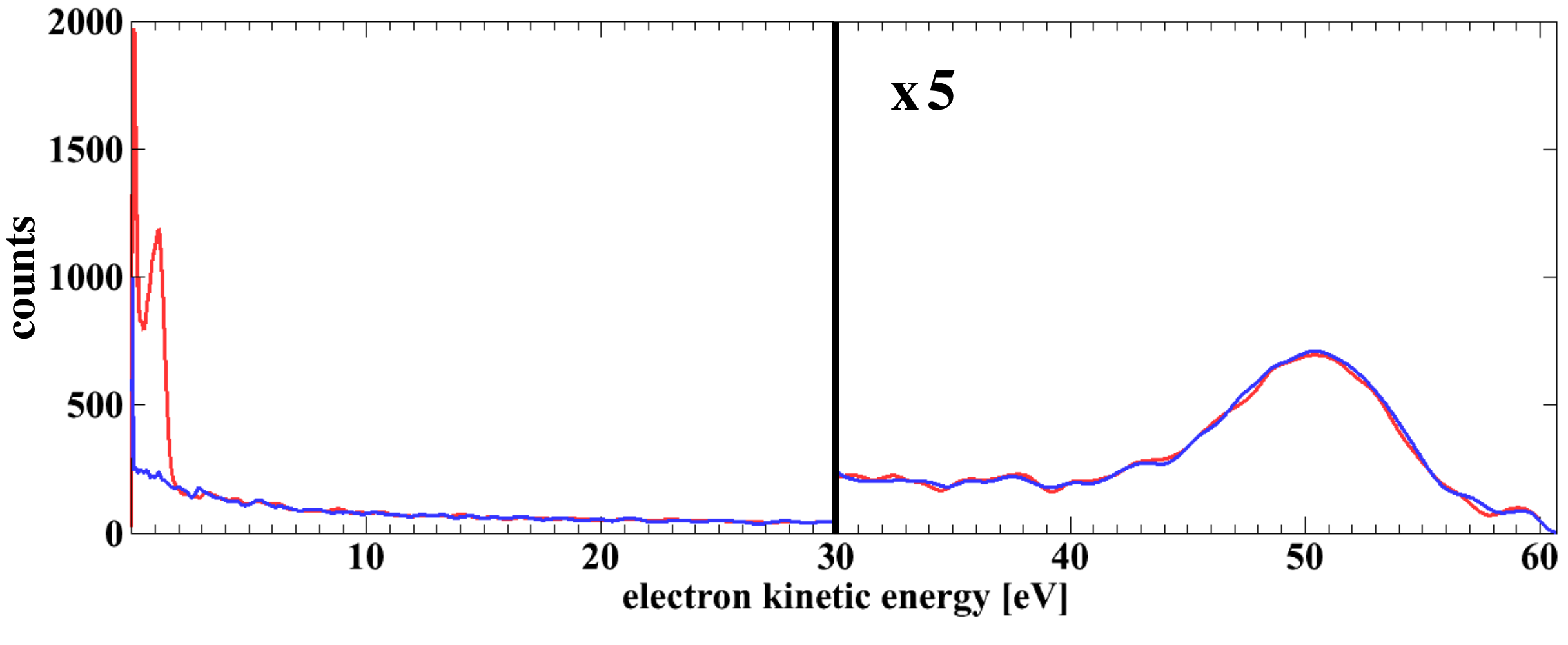}}
\caption{(a) Zoom-in on the central part of the electron difference image shown in Fig.~\ref{f_el_on_m_off}. (b) Electron energy spectrum recorded with (red) and without (blue) the YAG alignment laser, obtained from inversion of the detector images with pBasex. The spectrum on the right of the vertical bar is multiplied by a factor of 5.}
\end{figure}

Since the post-ionization of excited fragments should also result in the creation of additional electrons, we now investigate the difference between electron detector images recorded with and without the alignment laser, shown in Fig.~\ref{f_yag_el}, zoomed in to the central part of the detector. Clearly, two additional contributions of electrons with low energies emerge when the YAG is present. These can also be clearly identified in the electron spectrum  shown in Fig.~\ref{f_spectrum}. The two features are found to have maxima at electron energies of approximately 0.15 and 1.3\,eV, as calibrated with a measurement of the above-threshold-ionization in argon performed with the same spectrometer voltages. We note that the difference in kinetic energy between those two lines corresponds, within the uncertainties of our energy calibration, to the YAG photon energy of 1.17\,eV, which suggests that the two channels may result from n- and (n+1)-photon ionization of electronically excited molecules, molecular clusters, or fragments by the YAG pulse, although the exact origin is unclear to us at this point. In particular, it is surprising that two clear lines appear in the electron spectrum rather than a broad feature which one might expected if a series of close-lying Rydberg states was ionized.

Turning to the F($1s$)-photoelectron line at 51\,eV kinetic energy in Fig.~\ref{f_spectrum}, we notice that it is rather broad. This can be understood keeping in mind that the FEL pulses at the LCLS are created from self-amplified spontaneous emission~(SASE) and therefore have an intrinsic bandwidth of 0.2\,-\,1.0\,\%. \cite{emma_first_2010} This corresponds to a bandwidth of up to 7.4\,eV at an \xray~energy of 742\,eV, which cannot be reduced even when sorting on the shot-to-shot photon energy information.\footnote{Depending on the operation mode of the linear accelerator, there may also be systematic shifts in photon energy between different 30\,Hz sub-sets of the full 120\,Hz repetition rate, as we noticed in some of our data recorded in 2011.}

\begin{figure} [bt]
\centering
\includegraphics[width = 0.485\textwidth]{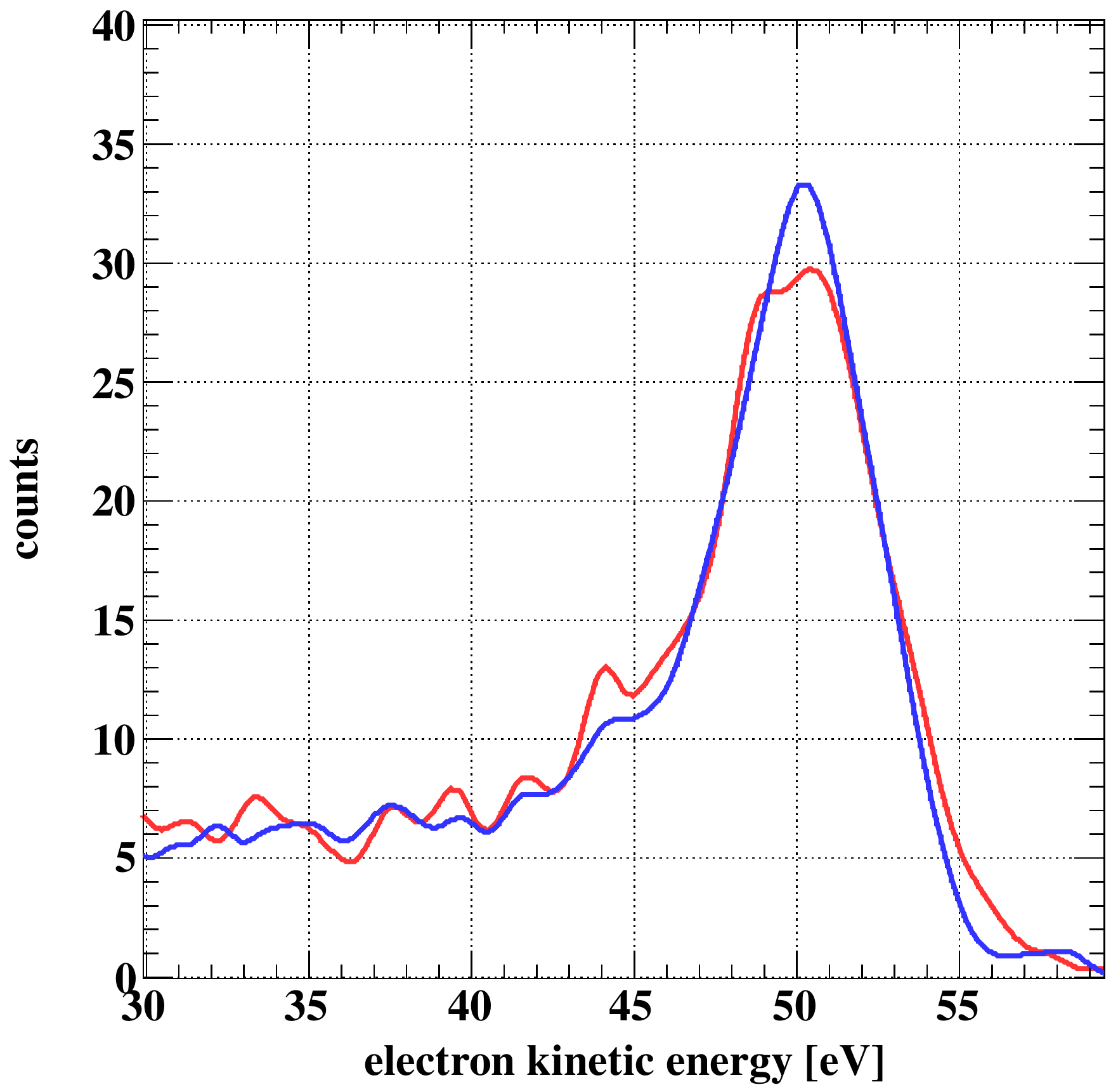}
\caption{\label{f_spectrum_90}Photoelectron energy spectra recorded with (red) and without (blue) YAG alignment laser pulse in a cone with an opening angle of 10$^\circ$ around the laser polarization direction.}
\end{figure}

Focusing on the photoline in more detail, we can investigate two-color effects on the inner-shell photoelectrons. While the ponderomotive broadening of the photoline due to the field of the YAG pulses is negligible for the given YAG pulse intensity, another possible direct influence of the YAG laser pulses on the photoelectrons is the formation of sidebands.\cite{meyer_two-colour_2010, kazansky_effect_2012} When the \xray~and alignment laser pulses are present at the same time, the photoelectron can absorb one or more YAG photons in addition to the \xray~photon in a process referred to as two-color above threshold ionization. Each YAG photon can increase or decrease the nominal electron kinetic energy by 1.17\,eV, resulting in a splitting of the photoline in multiple sub-lines, which is strongest for electron emission parallel to the YAG polarization direction. Given the bandwidth of the FEL pulses, the individual sidebands cannot be resolved in this photoelectron spectrum. Nevertheless, a slight broadening of the photoline recorded in the presence of the YAG pulse is observed when the energy spectrum is analyzed within 10$^\circ$ around the laser polarization direction, see Fig.~\ref{f_spectrum_90}, which may be caused by the formation of sidebands. This broadening and especially its angular dependence can also be seen more clearly in the detector difference image in Fig.~\ref{f_el_on_m_off}. However, for the analysis of the effects of molecular alignment on the photoelectron angular distributions described in section~3.3, we have assumed that the creation of sidebands does not significantly affect the photoelectron angular distribution as long as the photoelectron intensity is integrated over all sidebands.

\subsection{Effects of the pump laser}
\label{TiSa}

\begin{figure} [tb]
\centering
\subfigure{\includegraphics[width = 1.0\textwidth]{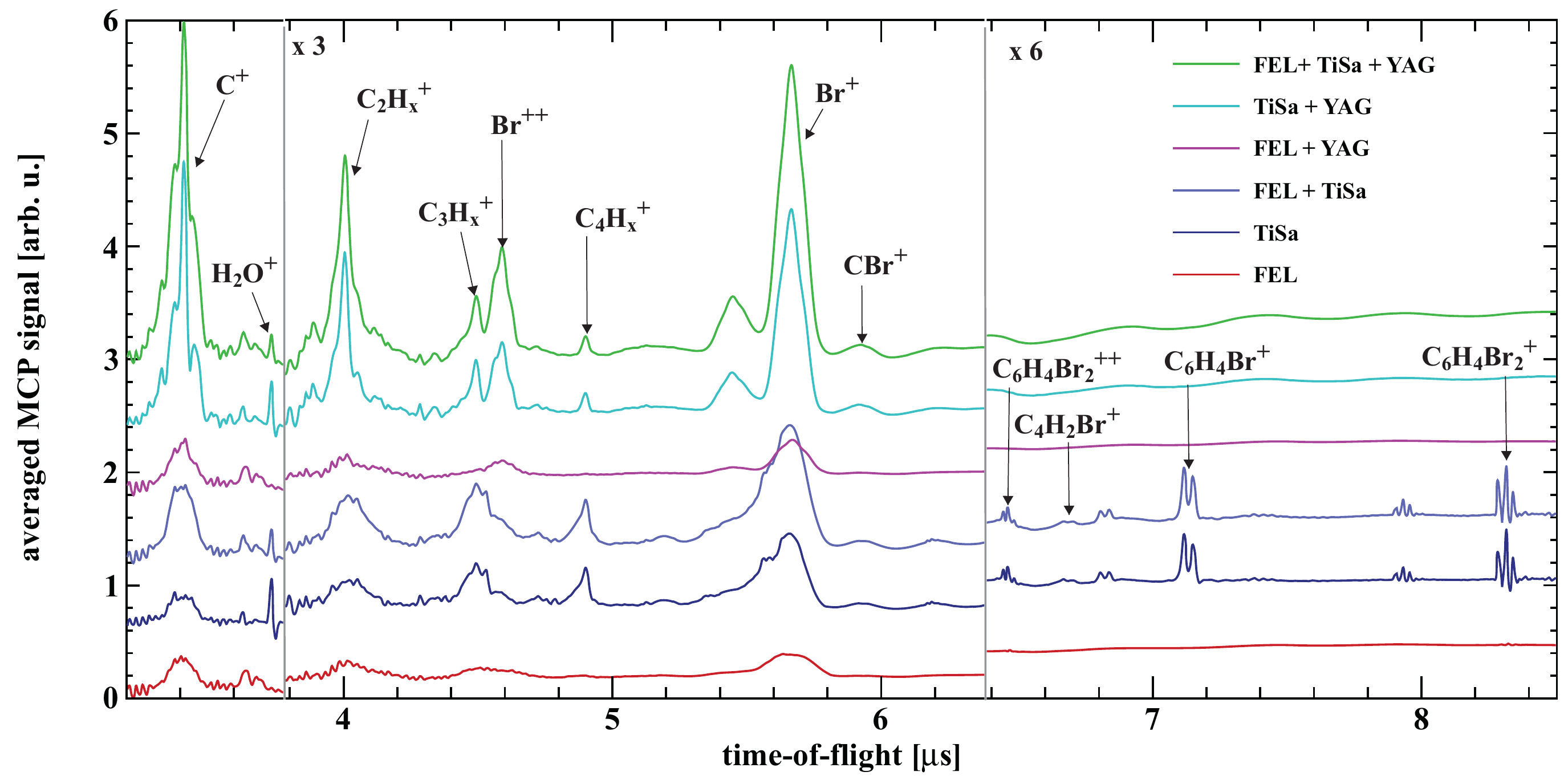}}
\caption{\label{f_DBB_tof}Ion time-of-flight spectra of DBB measured at the LCLS for a photon energy of 1570\,eV and different combinations of FEL, YAG, and TiSa pulses. For the cases with FEL and TiSa pulses present, the TiSa pulse arrives 0.5\,ps after the FEL pulse. The YAG pulse alone is non-ionizing. The polarization directions of FEL, YAG, and TiSa pulses are parallel to the detector plane. The traces beyond 3.8\,$\mu$s are scaled up by a factor of 3, the traces beyond 6.4\,$\mu$s are scaled up by a factor of 6. As in Fig.~\ref{f_yag_scan}, some small, additional peaks to the left of each main peak are due to secondary electrons created on the drift tube mesh.}
\end{figure}

Although a femtosecond TiSa laser was part of the experimental setup of the pFAB experiment and was used to optimize the molecular alignment, we did not perform a pump-probe experiment for lack of time, thus only static photoelectron angular distributions were investigated. 

In the earlier experiment on 1,4-dibromobenzene (DBB) molecules, a TiSa pulse was used to dissociate the molecules before they were ionized by the FEL pulse. The photoelectron angular distributions recorded in that experiment are described elsewhere.\cite{rolles_femtosecond_2014} Here, we concentrate on the influence of the three different light pulses on the molecular fragmentation as seen in the ion time-of-flight spectra shown in Fig.~\ref{f_DBB_tof}, which were recorded simultaneously to the electron images reported in \cite{rolles_femtosecond_2014}. Note that during the DBB experiment, a plate with a 0.5-mm wide slit perpendicular to the FEL beam propagation direction was placed inside of the spectrometer in order to only accept ions that were created in the center of the spectrometer. This limits significantly the angular acceptance for energetic fragment ions, and these spectra therefore only allow a qualitative investigation of the fragmentation. Furthermore, the first 3\,$\mu$s of the spectrum are heavily disturbed by high-frequency pickup from the high-voltage switching on the electron detector on the opposite side of the spectrometer, thus the spectra are only shown for mass-to-charge ratios beyond C$^+$.

As for the case of pFAB discussed above,  inner-shell ionization with an \xray~photon alone (red trace), here at a photon energy of 1570\,eV, i.e.\,roughly 20\,eV above the Br($2p_{3/2}$) threshold but still below the Br($2p_{1/2}$) threshold, creates various charged fragments, mostly \Bion~as well as C$_3$H$_x^+$, C$_2$H$_x^+$, and C$^+$ ions. A very small amount of parent ions is also created, either due to valence ionization or due to fluorescent decay of the core-hole. The parent ion peak, which is almost invisible in the FEL spectrum, is very sharp though, so no indications for the formation of clusters in the supersonic expansion are observed in this data.

The TiSa pulse alone (dark blue trace) creates singly charged parent ions (with a triple structure due to the bromine isotopes) along with a variety of other singly charged fragments. A small amount of doubly parent ions occurs as well, but most of the doubly charged molecules decay further in smaller fragments, most prominently \Bion. When a TiSa pulse interacts with the molecules after the FEL pulse ionized them (lighter blue trace), only small changes can be seen in the ion TOF spectrum as compared to the spectrum recorded with only the TiSa pulse present. This is understandable since the focus of the TiSa beam was chosen larger than the focus of the FEL beam to ensure that all molecules probed by the \xray s were also in the focus of the pump laser. Furthermore, the cross section for ionization with the TiSa at this intensity is higher than the cross section for ionization with the X-rays. Therefore, significantly more molecules are ionized by the TiSa laser pulse alone and the spectrum is thus dominated by these ions.

When both \xray~and YAG pulses are present (purple trace), the spectrum does not change significantly from the spectrum observed for \xray~pulses alone, although a small increase in the yield of certain ions can be observed. This is very different from what was found in the pFAB data in the previous section. We note, however, that a significant amount of clusters was present in the pFAB experiment, which did not seem to be the case for DBB.

When the strong-field ionization by the TiSa is combined with the pulses from the YAG laser (cyan trace), the changes in the ion TOF spectrum are more dramatic. All fragments heavier then \Bion~disappear, while almost all other peaks are strongly enhanced, indicating that the combination of YAG and TiSa pulses ionizes more strongly than the TiSa pulse alone. We tentatively explain this as the effect of dissociation, single- or multi-photon ionization of excited molecular fragments, which are created by the TiSa pulse, by the YAG pulse. When the \xray~pulse is added to the TiSa and YAG pulses (green trace), the spectrum is again dominated by the fragmentation induced by TiSa and YAG pulses because of the larger focus of TiSa and YAG beams as compared to the X-ray beam and higher cross sections for ionization by the TiSa pulses. 

Summarizing our findings for the DBB molecules and the discussion of the effects of the YAG pulse in the pFAB data in the previous section, we can conclude that the field of the YAG laser pulse apparently has a strong influence on the ionization and fragmentation dynamics. At this point, we have no direct evidence that this changes the photoelectron angular distributions, but it certainly gives reason to suspect that the molecular dynamics initiated by a femtosecond pump pulse may be influenced by the presence of the strong field of the YAG pulse. A possibility to circumvent this effect could be to use either impulsive, "field-free" alignment or electron-ion coincidence techniques to align or orient the molecules in space, but as we briefly discuss in the following section, these techniques also have practical limitations.

Concerning the "pump" process, we note that Coulomb explosion by a strong 800-nm TiSa pulse was used here mostly as a proof-of-principle. In order to selectively trigger photochemical reactions, a single-photon transition to a resonant excitation, ideally by a non-ionizing laser pulse, would, in many cases, be more appropriate.

\section{Conclusions and Outlook}

In this paper, along with our previous publications on this subject\cite{krasniqi_imaging_2010,boll_femtosecond_2013, rolles_femtosecond_2014}, we have reported the current status of our efforts to perform femtosecond time-resolved photoelectron diffraction experiments on gas-phase molecules in a pump-probe setup combining optical lasers and an \xray~Free-Electron Laser. We have presented results of two photoelectron and ion imaging experiments on laser-aligned 1-ethynyl-4-fluorobenzene (C$_8$H$_5$F) and 1,4-dibromobenzene (C$_6$H$_4$Br$_2$) molecules conducted at the LCLS and compared some of the results with photoelectron-photoion coincidence data recorded at the PETRA synchrotron radiation facility. We have also discussed the contribution of molecular clusters to our experimental data on 1-ethynyl-4-fluorobenzene as well as the influence of the nanosecond alignment laser pulse and the femtosecond pump laser pulse on the photoelectrons and on the molecular fragmentation.

Our results demonstrate that by combining a strong nanosecond YAG laser pulse with the FEL pulse, it is possible to perform photoionization experiments on adiabatically laser-aligned and mixed-field oriented polyatomic molecules. The corresponding photoelectron angular distributions show a clear dependence on the photoelectron kinetic energy,\cite{boll_femtosecond_2013} on the alignment direction of the molecular axis,\cite{rolles_femtosecond_2014} and on the molecular orientation. While our interpretation was, so far, mostly based on comparison to density function theory calculations \cite{boll_femtosecond_2013, rolles_femtosecond_2014} our long-term goal is to link the observed patterns directly to the molecular structure by applying the concepts of photoelectron diffraction and holography.\cite{krasniqi_imaging_2010} 

Time-resolved photoelectron diffraction and holography has the potential to image the geometric structure of gas-phase molecules with few-femtosecond temporal and sub-\ang~spatial resolution, and offers a complementary approach to time-resolved \xray~and electron diffraction. Using electrons as opposed to \xray s for diffraction has the advantage of much higher elastic scattering cross sections, which is particularly important for targets containing lighter atoms such as carbon, nitrogen, or oxygen, which do not scatter X-rays efficiently. Using photoelectrons instead of an electron beam has the additional benefit of avoiding the problem of velocity mismatch in laser-pump electron-probe experiments on gas-phase targets.

A disadvantage of photoelectron diffraction, however, is the more complicated description of the initial photoelectron wave, which, contrary to the case of \xray~and electron diffraction, does not fulfill the plane-wave approximation. In the case of photoionization of an inner-shell $s$-orbital, the initial, unscattered photoelectron wave can be described, to a good approximation, by a pure $p$-wave, while for photoionization of orbitals with an angular momentum quantum number $l \neq 0$, the interference of the $l-1$ and $l+1$ partial waves already complicates the description of the unscattered photoelectron wave.  Furthermore, the interpretation of the final photoelectron angular distributions in terms of scattering is particularly challenging for low-energy electrons, where the molecular potential can no longer be approximated by the sum of atomic potentials and where multiple scattering can be a significant contribution. Nevertheless, we are convinced that detailed insight into changes of the molecular structure during photochemical reactions can be gained from studying photoelectron angular distributions even in these more difficult cases. Our goal is therefore to establish the photoelectron diffraction concept for gas-phase molecules while developing the experimental tools to perform these experiments in a femtosecond pump-probe setup.

In the experiments we have reported so far, the degree of molecular alignment that we achieved was sufficiently high to observe alignment dependent effects when considering the difference between the photoelectron angular distributions of aligned and unaligned molecules. In order to obtain more direct information on the molecular structure, e.g. by holographic reconstruction, \cite{krasniqi_imaging_2010} a considerably higher degree of alignment is necessary. Such high degrees of alignment up to $\langle \cos^2 \theta_{2D}\rangle$\,=\,0.97 have been achieved for iodobenzene molecules using adiabatic laser alignment in a laboratory setup. \cite{holmegaard_laser-induced_2009} Since the presence of the strong laser field used for adiabatic alignment may cause unwanted effects in pump-probe experiments, as discussed in section 3.4 and 3.5, an option to circumvent these effects could be to use field-free alignment techniques. However, these have, so far, not been able to obtain as high degrees of alignment as adiabatic techniques. 

For suitable classes of molecules, an alternative way to determine the molecular alignment and orientation very precisely is by means of electron-ion coincidence techniques. This may become a competetive option once higher repetition-rate FEL sources such as the European XFEL are available. However, as our discussion of the fragmentation of pFAB molecules after inner-shell ionization has shown, this may also be challenging for polyatomic molecules, where complicated fragmentation channels and the occurrence of a large number of possible fragments can make it difficult or impossible to find a fragmentation channel that is suitable to define one or several molecular axes.

Finally, the experiments reported here still lack the necessary temporal resolution to resolved dynamics on the order of 100\,fs or below, but the use of X-ray optical cross-correlation techniques \cite{bionta_spectral_2011, schorb_x-rayoptical_2012, harmand_achieving_2013} was shown to improve this dramatically. With the lessons learned from our previous experiments, we therefore believe that there is a clear avenue towards a time-resolved photoelectron diffraction experiment that would be able to image the molecular structure during an isomerization reaction or close to transition states in a photochemical reaction by measuring photoelectron angular distributions as suggested, e.g., in \cite{boll_femtosecond_2013}.

Part of this research was carried out at the Linac Coherent Light Source (LCLS) at the SLAC National Accelerator Laboratory. LCLS is an Office of Science User Facility operated for the U.S. Department of Energy Office of Science by Stanford University. Additional measurements were performed at the PETRA P04 beamline at DESY. We acknowledge the Max Planck Society for funding the development and operation of the CAMP instrument within the ASG at CFEL. D.R. acknowledges support from the Helmholtz Gemeinschaft through the Young Investigator Program. L.C., S.D., and H.S. acknowledge support from the Carlsberg Foundation. J.K. acknowledges support from the excellence cluster “The Hamburg Centre for Ultrafast Imaging - Structure, Dynamics and Control of Matter at the Atomic Scale” of the Deutsche Forschungsgemeinschaft. A.Ro. and M.V acknowledge the research program of the "Stichting voor Fundamenteel Onderzoek der Materie", which is financially supported by the "Nederlandse organisatie voor Wetenschappelijk Onderzoek". P.J. acknowledges support from the Swedish Research Council and the Swedish Foundation for Strategic Research. A.Ru. acknowledges support from the Office of Basic Energy Sciences, US Department of Energy. S.Te. and A.S. acknowledge support through SFB 755 Nanoscale photonic imaging. We are grateful to the entire LCLS and PETRA staff for their support and hospitality during the beamtimes.

\footnotesize{
\bibliography{faraday_arxiv} 
\bibliographystyle{rsc}
}

\end{document}